\documentclass[aps,preprint,epsfig,rotate]{revtex4}
\usepackage{graphicx}
\usepackage{bm}
\usepackage{epsfig}

\input epsf
\begin{document}
\title{Compact and accurate variational wave functions of three-electron
       atomic systems constructed from semi-exponential radial basis
       functions}

 \author{Alexei M. Frolov}
 \email[E--mail address: ]{afrolov@uwo.ca}

\affiliation{Department of Chemistry\\
 University of Western Ontario, London, Ontario N6H 5B7, Canada}

\date{\today}

\begin{abstract}

The semi-exponential basis set of radial functions (A.M. Frolov, Physics
Letters A {\bf 374}, 2361 (2010)) is used for variational computations of
bound states in three-electron atomic systems. It appears that
semi-exponential basis set has a substantially greater potential for
accurate variational computations of bound states in three-electron atomic
systems than it was originally anticipated. In particular, the 40-term
Larson's wave function improved with the use of semi-exponential radial
basis functions now produces the total energy \linebreak -7.47805413551
$a.u.$ for the ground $1^2S-$state in the ${}^{\infty}$Li atom (only one
spin function $\chi_1 = \alpha \beta \alpha - \beta \alpha \alpha$ was used
in these calculations). This variational energy is very close to the exact
ground state energy of the ${}^{\infty}$Li atom and it substantially lower
than the total energy obtained with the original Larson's 40-term wave
function (-7.477944869 $a.u.$).

PACS number(s): 03.65.Ge, 31.15.ac and 31.15.xf
\end{abstract}
\maketitle

\section{Introduction}\label{intro}

In this study we perform variational calculations of bound states in
three-electron atomic systems. An advanced basis set of semi-exponential
radial functions \cite{Fro2010} is extensively used in our calculations. The
main goal is to solve the non-relativistic Schr\"{o}dinger equation $H \Psi
= E \Psi$, where $E < 0$ and bound state wave function $\Psi$ has the unit
norm. The general non-relativistic Hamiltonian $H$ of the three-electron
atomic problem is (see, e.g., \cite{LLQ})
\begin{eqnarray}
 H = -\frac{\hbar^2}{2 m_e} \Bigl[\nabla^2_1 + \nabla^2_2 +
      \nabla^2_3 + \frac{m_e}{M} \nabla^2_4 \Bigr]
     - \frac{Q e^2}{r_{14}} - \frac{Q e^2}{r_{24}}
     - \frac{Q e^2}{r_{34}} + \frac{e^2}{r_{12}} + \frac{e^2}{r_{13}}
     + \frac{e^2}{r_{23}} \label{Hamil}
\end{eqnarray}
where $\hbar = \frac{h}{2 \pi}$ is the reduced Planck constant, $m_e$ is the
electron mass and $e$ is the electric charge of electron. In this equation
and everywhere below in this study the subscripts 1, 2, 3 designate the
three electrons $e^-$, while the subscript 4 denotes the heavy nucleus with
the mass $M$ ($M \gg m_e$) and positive electric (nuclear) charge $Q e$. The
notations $r_{ij} = \mid {\bf r}_i - {\bf r}_j \mid = r_{ji}$ stand for the
six interparticle distances (= relative coordinates) defined in an arbitrary
four-body system and ${\bf r}_i$ ($i$ = 1, 2, 3, 4) are the Cartesian
coordinates of the four point particles. In Eq.(\ref{Hamil}) and everywhere
below in this work we shall assume that $(ij)$ = $(ji)$ = (12), (13), (14),
(23), (24), (34). In fact, below only atomic units $\hbar = 1, \mid e \mid =
1, m_e = 1$ are employed. In these units the explicit form of the
Hamiltonian $H$, Eq.(\ref{Hamil}), is significantly simplified.

The main attention in this work is focused on numerical calculations of the
ground (doublet) $1^2S(L = 0)-$state (or $1^2S_{\frac12}(L = 0)-$state) of
the three-electron Li atom with the infinitely heavy nucleus, i.e. the
${}^{\infty}$Li atom. Considerations of other three-electron atoms, ions
and various positron containing atomic systems (e.g., HPs) can be performed
absolutely analogously (for more detail, see \cite{Fro2010}) and, therefore,
these systems will not be considered here.

The problem of highly accurate calculations of the bound states in
three-electron atomic systems has attracted a very significant attention.
First calculations of the Li atom with the truly correlated wave functions
were performed in 1936 \cite{JC}. A brief reviews of such calculations can
be found in \cite{McW1} (earlier works) and \cite{King1} (references up to
1997 are mentioned). The current bibliography on this subject includes
almost one thousand references and rapidly increases. At this moment the
classical Hylleraas method (see, e.g., \cite{King1}) is only the method used
to construct highly accurate wave functions for bound states in
three-electron atomic systems. In this method to produce the highly accurate
wave functions, e.g., for the ground (doublet) $1^2S$-state in the Li atom
one needs to use many thousands Hylleraas basis functions. The use of
extremely large basis sets is very inconvenient in many actual cases, since
it produces a number of computational problems. It is clear that the
Hylleraas method cannot be used to construct compact and accurate
variational wave functions for three-electron systems. Indeed, it contains
essentially no control parameters which can be optimized by increasing the
overall efficiency of the method.

An alternative approach to variational bound state calculations in
three-electron atomic systems was proposed in \cite{Fro2010}. This approach
is based on the use of semi-exponential variational wave functions
\cite{Fro2010} and it allows one to construct very compact and accurate
variational wave functions in arbitrary three-electron atomic system. Each
of the semi-exponential basis functions depend upon all six interparticle
coordinates $r_{12}, r_{13}, \ldots, r_{34}$ \cite{Fro2010}. A very high
efficiency of this new approach in actual applications and its superiority
over the classical Hylleraas expansion was demonstrated in \cite{Fro2010}.
As follows from the results of this study the semi-exponential basis set has
a substantially greater potential for highly accurate variational
computations of bound states in three-electron atomic systems than it was
anticipated earlier \cite{Fro2010}.

\section{Variational wave function}

The variational wave function of the doublet $S(L = 0)-$states of the
three-electron Li atom is written in the following general form
\begin{eqnarray}
 \Psi_{L=0} = \psi_{L=0}(A; \bigl\{ r_{ij} \bigr\}) (\alpha \beta \alpha
 - \beta \alpha \alpha) + \phi_{L=0}(B; \bigl\{ r_{ij} \bigr\}) (2 \alpha
 \alpha \beta  - \beta \alpha \alpha - \alpha \beta \alpha) \label{psi}
\end{eqnarray}
where $\psi_{L=0}(A; \bigl\{ r_{ij} \bigr\})$ and $\phi_{L=0}(B; \bigl\{
r_{ij} \bigr\})$ are the two independent spatial parts (= radial parts) of
the total wave function. Each of these two radial functions is, in fact, a
radial factor (for states with $L = 0$) in front of the corresponding
three-electron spin functions $\chi_1 = \alpha \beta \alpha - \beta \alpha
\alpha$ and $\chi_2 = 2 \alpha \alpha \beta  - \beta \alpha \alpha - \alpha
\beta \alpha$. Here the notations $\alpha$ and $\beta$ are the one-electron
spin-up and spin-down functions, respectively (see, e.g., \cite{Dir}). The
notations $A$ and $B$ in Eq.(\ref{psi}) mean that the two sets of non-linear
parameters associated with radial functions $\psi$ and $\phi$ can be
optimized independently. In the general case, each of the radial basis
functions explicitly depends upon all six interparticle (relative)
coordinates $r_{12}, r_{13}, r_{23}, r_{14}, r_{24}, r_{34}$. It is clear
that in actual variational calculations only one spin function, e.g., the
$\chi_1$ function, can be used.  Another useful trick (so-called
`doubling') is based on the use of the same set of non-linear parameters in
the two radial parts in Eq.(\ref{psi}).

In our earlier work \cite{Fro2010} we have introduced an advanced set of
radial basis functions for three-electron atomic calculations. In
\cite{Fro2010} this set was called the semi-exponential basis set. In
general, the semi-exponential variational expansion of the radial function
$\psi_{L=0}(A; \bigl\{ r_{ij} \bigr\})$ is written in the form
\begin{eqnarray}
 \psi_{L=0}(A; \bigl\{ r_{ij} \bigr\}) = \sum^N_{k=1} C_k r^{n_1(k)}_{23}
 r^{n_2(k)}_{13} r^{n_3(k)}_{12} r^{m_1(k)}_{14} r^{m_2(k)}_{24}
 r^{m_3(k)}_{34} exp(-\alpha_{k} r_{14} -\beta_{k} r_{24} -\gamma_{k}
 r_{34}) \label{semexp}
\end{eqnarray}
where $\alpha_k, \beta_k, \gamma_k$ ($k = 1, 2, \ldots, N$) are the varied
non-linear parameters. The presence of the varied non-linear parameters in
Eq.(\ref{semexp}) is the main and very important difference with the
traditional Hylleraas variational expansion (see, e.g., \cite{Lars}) for
which in Eq.(\ref{semexp}) we always have $\alpha_1 = \ldots = \alpha_N,
\beta_1 = \ldots = \beta_N$ and $\gamma_1 = \ldots = \gamma_N$. Note that
all matrix elements of the Hamiltonian, Eq.(\ref{Hamil}), and overlap matrix
needed in computations with the use of the semi-exponential basis,
Eq.(\ref{semexp}), contain the same three-electron integrals which arise for
the usual Hylleraas expansion (for more detail, see \cite{Fro2010}). In
other words, numerical calculation of all matrix elements with our
semi-exponential functions is no more difficult problem, than for the
traditional Hylleraas radial functions. This also simplifies numerical
computation of the bound state properties (i.e. expectation values) in the
semi-exponential basis set. Our algorithms used in calculations of all
required matrix elements is based on the Perkins formula for three-electron
integrals \cite{Per} in relative coordinates. Note also that all
calculations in this work have been performed with the use of standard
quadruple precision accuracy (30 decimal digits per computer word).

In actual atomic systems any many-electron wave function must be completely
antisymmetric upon all electron variables, i.e. upon all electron spatial
and spin variables. For three-electron atomic wave function this requirement
is written in the form ${\hat{\cal A}}_{123} \Psi(1,2,3) = - \Psi(1,2,3)$,
where $\Psi$ is given by Eq.(\ref{psi}) and $\hat{{\cal A}}_e$ is the
three-particle (= electron) antisymmetrizer ${\hat{\cal A}}_e = \hat{e} -
\hat{P}_{12} - \hat{P}_{13} - \hat{P}_{23} + \hat{P}_{123} + \hat{P}_{132}$.
Here $\hat{e}$ is the identity permutation, while $\hat{P}_{ij}$ is the
permutation of the $i$-th and $j$-th particles. Analogously, the operator
$\hat{P}_{ijk}$ is the permutation of the $i$-th, $j$-th and $k$-th
particles. In actual computations antisymmetrization of the total wave
function is reduced to the proper antisymmetrization of corresponding matrix
elements (for more detail, see, e.g., \cite{Fro2010}). Each of these matrix
elements is written in the form $\langle \Psi \mid \hat{O} \mid \Psi
\rangle$, where $\hat{O}$ is an arbitrary spin-independent quantum operator
which is truly symmetric upon all interparticle permutations. The wave
function $\Psi$, Eq.(\ref{psi}), contains the two different radial parts
$\psi$ and $\phi$. By performing the integration over all spin coordinates
from here one finds the four spatial projectors ${\cal P}_{\psi\psi}, {\cal
P}_{\psi\phi} = {\cal P}_{\phi\psi}$ and ${\cal P}_{\phi\phi}$ presented in
\cite{Fro2010}. In fact, the explicit form of the ${\cal P}_{\psi\phi}$ and
${\cal P}_{\phi\psi}$ projectors given in \cite{Fro2010} must be corrected
(there is an obvious misprint in the formulas given in \cite{Fro2010})
\begin{eqnarray}
 {\cal P}_{\psi\phi} &=& \frac12 \Bigl( \hat{P}_{13} - \hat{P}_{23} +
 \hat{P}_{123} - \hat{P}_{132} \Bigr) \\
 {\cal P}_{\phi\psi} &=& \frac12 \Bigl( \hat{P}_{13} - \hat{P}_{23} +
  \hat{P}_{123} - \hat{P}_{132} \Bigr)
\end{eqnarray}
For an arbitrary truly symmetric spin-independent operator $\hat{O}$ each
of these four projectors produces matrix elements $\langle \Psi \mid \hat{O}
\mid \Psi \rangle$ of the correct permutation symmetry (for doublet states)
between all three electrons. The explicit formulas for all matrix elements
obtained with the radial basis functions, Eq.(\ref{semexp}), and for
three-electron integrals needed in calculations can be found in
\cite{Fro2010}.

\section{Calculations}

Let us apply the semi-exponential variational expansion, Eq.(\ref{semexp}),
to numerical calculations of the ground $1^2S$-state in the three-electron
${}^{\infty}$Li atom. In this study we consider the two variational wave
functions: (a) the wave function which contains 28 radial basis functions,
Eq.(\ref{semexp}), and (b) the wave function which includes 40 radial basis
functions, Eq.(\ref{semexp}). The results (in atomic units) obtained with
these two trial wave functions can be found in Table I. Tables II and III
contain the corresponding radial basis functions, Eq.(\ref{semexp}), i.e.
the powers $n_1(k), n_2(k), n_3(k), m_1(k), m_2(k), m_3(k)$ of six radial
variables $r_{12}, r_{13}, r_{23}, r_{14}, r_{24}, r_{34}$ and optimized
non-linear parameters $\alpha_k, \beta_k, \gamma_k$. As follows from Table I
our variational energies obtained for the ground $1^2S$-state in the
${}^{\infty}$Li atom with the use of semi-exponential variational expansion,
Eq.(\ref{psi}), are substantially lower than the corresponding energies
determined for this state with the same Hylleraas wave function \cite{Lars}.
Note also that the non-linear parameters used in our method (in
Eq.(\ref{semexp})) are constantly varied. Therefore, it is hard to say that
the total energies obtained in some calculations are `final'. Formally,
based on the known convergence rate(s) for our data and by using a few
extrapolation procedures we can approximately evaluate the limits to which
our variational energies will converge, if we could perform an infinite
number of variations for the non-linear parameters in Eq.(\ref{semexp}).
Such limits for the total energies are shown in the fourth column of Table
I. These values indicate that, e.g., our 40-term variational wave function
can produce, in principle, very accurate variational energies, if the
optimization of non-linear parameters in Eq.(\ref{semexp}) will continue.

In \cite{Lars} Larsson proposed a simple (but useful!) trick which allows
one to increase the overall accuracy of the trial (doublet) wave function.
Later this trick was called `doubling' of the wave function and it was used
practically in all calculations of the bound doublet states in
three-electron atomic systems. The idea of doubling is simple and
transparent. If we already know the radial function constructed for one spin
configuration, e.g., for $\chi_1 = \alpha \beta \alpha - \beta \alpha
\alpha$ from Eq.(\ref{psi}), then we can use exactly the same radial basis
function for another spin configuration $\chi_2 = 2 \alpha \alpha \beta -
\beta \alpha \alpha - \alpha \beta \alpha$. Formally, it doubles the total
number of basis functions in the trial wave function. According to the
variational principle the total variational energy can only decrease during
such a procedure. The problem of linear dependence of basis functions is
avoided in this procedure, since the two spin functions $\chi_1 = \alpha
\beta \alpha - \beta \alpha \alpha$ and $\chi_2 = 2 \alpha \alpha \beta -
\beta \alpha \alpha - \alpha \beta \alpha$ are independent of each other. In
fact, for Hylleraas variational expansion the `doubling' does not work
properly, since there are obvious linear dependencies between different
radial basis functions in those cases when some non-linear parameters
coincide with each other (for more detail, see \cite{Lars}). In
semi-exponential variational expansion Eq.(\ref{semexp}) all optimized
non-linear parameters are independent of each other. Therefore, the
coincidence of the pre-exponential factors in Eq.(\ref{semexp}) is not
crucial and does not mean that such basis functions are linearly dependent.
This drastically simplifies the actual `doubling' for Eq.(\ref{semexp}). The
energies obtained with the use of `doubling' of our variational wave
functions can be found in the fifth column of Table I. It is clear that the
non-linear parameters (or parameters in the exponents in Eq.(\ref{semexp}))
from the second part of the total wave function are not optimal, i.e. for
all terms which contain basis functions with numbers $i \ge 41$ the
non-linear parameters are not optimal. These 120 (= 40 $\times 3$)
non-linear parameters can be re-optimized and this drastically improves the
overall quality of the total wave function. For instance, approximate
one-step re-optimization of the last 40 non-linear parameters in the wave
function gives the ground state energy -7.47805456511 $a.u.$, which is much
better than the `doubling' energy from Table I.

As follows from Table I the doubling is not an effective approach for our
trial wave functions with the carefully optimized non-linear parameters.
However, we can modify the original idea of doubling into something new
which is substantially more effective in actual computations. To illustrate
one of such modifications, let us assume that we have constructed 40-term
variational wave function, Eq.(\ref{semexp}) which contains 40 $\times$ 3 =
120 carefully optimized non-linear parameters $\alpha_1, \beta_1, \gamma_1,
\ldots, \alpha_{40}, \beta_{40}, \gamma_{40}$. At the second step of our
procedure we can add forty additional basis functions with the same
pre-exponential factors $r^{n_1(k)}_{23} r^{n_2(k)}_{13} r^{n_3(k)}_{12}
r^{m_1(k)}_{14} r^{m_2(k)}_{24} r^{m_3(k)}_{34}$, but slightly different
exponents in Eq.(\ref{semexp}). In reality, these new exponents have been
chosen quasi-randomly from three different intervals, e.g.,
\begin{eqnarray}
 \alpha_{i + 40} = \alpha_{i} + 0.0057 \cdot \Bigl<\Bigl< \frac{i (i + 1)
 \sqrt{2}}{2} \Bigr>\Bigr> \nonumber \\
 \beta_{i + 40} = \beta_{i} + 0.0063 \cdot \Bigl<\Bigl< \frac{i (i + 1)
 \sqrt{3}}{2} \Bigr>\Bigr> \nonumber \\
 \gamma_{i + 40} = \gamma_{i} + 0.0049 \cdot \Bigl<\Bigl< \frac{i (i + 1)
 \sqrt{5}}{2} \Bigr>\Bigr> \nonumber
\end{eqnarray}
where $i$ = 1, 2, $\ldots$, 40 and $\Bigl<\Bigl< x \Bigr>\Bigr>$ designates
the fractional part of the real number $x$. Small deviations of these new
exponents from the known `optimal' values (i.e. from $\alpha_i, \beta_i,
\gamma_i$, where $1 \le i \le 40$) produce the extended wave function of
`almost optimal' quality. On the other hand, even these small differences
between exponents allows one to avoid a linear dependence between basis
vectors in Eq.(\ref{semexp}). Obviously, this procedure can be repeated a
number of times. This allows one to construct very accurate trial wave
functions which contain not only 80, but 400, 800 and even 2000 basis
functions with almost `optimal' non-linear parameters.

It is very interesting to perform variational calculations of the ground
state of the ${}^{\infty}$Li atom with the use of 60-term wave function
constructed from the analogous 60-term wave Larsson's wave function
\cite{Lars}. The variational total energy obtained in \cite{Lars} with that
wave function was -7.47801035965 $a.u.$ Our 60-term trial wave function with
one spin function $\chi_1$ constructed in \cite{Fro2010} corresponds to the
substantially lower total energy -7.478057561 $a.u.$ After publication of
\cite{Fro2010} we have started re-optimization of the non-linear parameters
in our 60-term trial wave function. The current total energy is
-7.47805637319 $a.u.$ (only one spin function $\chi_1$ is used in our
calculations). This energy is above the value from \cite{Fro2010}. However,
our current total energy rapidly decreases with almost constant rate
$\approx 2.5 \cdot 10^{-7}$ $a.u.$ per cycle, i.e. per one variation of all
180 (= 60 $\times$ 3) non-linear parameters in the trial wave function. We
expect that after an infinite number of variations of the non-linear
parameters the total energy of the ${}^{\infty}$Li atom obtained with our
60-term radial function and one spin function will converge to the value
-7.4780603(3) $a.u.$ which is very close to the actual ground state energy.
It will be an outstanding result to obtain the value lower than -7.4780602
$a.u.$ for the total energy of the ground $1^2S-$state in the
${}^{\infty}$Li atom by using only 60-term variational wave function.

\section{Conclusion}

The semi-exponential variational expansion \cite{Fro2010} is applied for
bound state calculations of three-electron atomic systems. It is shown that
this variational expansion allows one to construct very compact and accurate
variational wave functions for three-electron atomic systems. Currently, the
use of semi-exponential radial basis functions is the only way to produce
compact and accurate wave functions for three-electron atomic systems. The
total energies obtained in this study for the ground $1^2S$-state of the
${}^{\infty}$Li atom (see Table I) are more accurate than our earlier
results from \cite{Fro2010} and substantially more accurate than the
original Larsson's wave function \cite{Lars} with the same number of terms.
It indicates clearly that our semi-exponential variational expansion,
Eq.(\ref{semexp}), has a substantially greater potential for variational
bound state calculations in three-electron atomic systems than we have
anticipated originally \cite{Fro2010}. Currently, we continue the process of
numerical optimization of the non-linear parameters in our trial wave
functions constructed with the use of semi-exponential variational
expansion. Note that the choice of Larsson's wave function(s) as the first
approximation to the semi-exponential variational expansion is not crucial
for our method. Many other (different) choices are also possible.

The semi-exponential variational expansion is also applied for accurate
computations of various atomic and quasi-atomic three-electron systems. In
particular, we have made numerous improvements in our original computer code
\cite{Fro2010}. It is clear that the semi-exponential basis can be used to
construct variational wave functions in the four-electron atomic problems.
In particular, the semi-exponential variational expansion can be used to
obtain very compact and accurate wave functions of the ground singlet
$1^1S-$state and triplet $2^3S-$states of the Be atom and Be-like ions (for
more details, see \cite{HaSi}, \cite{FrWa} and reference therein). It is
clear that our method can also be used for rotationally excited states in
atomic systems, i.e. for states with $L \ge 1$, where $L$ is the electron
angular momentum.

\newpage
%
%
   \begin{table}
    \caption{The total energies $E$ (in atomic units) of the $1^2S(L =
             0)-$state in the ${}^{\infty}$Li atom. $N$ designates the
             number of basis functions used.}
      \begin{center}
      \begin{tabular}{lllll}
         \hline \hline 
 $N$ & $E$(Ref.[4]) & $E$(Eq.(3)) & $E^{a}$(Eq.(3)) & $E$(Eq.(3), doubling) \\
     \hline
 28 & -7.477885105 & -7.47803638005 & -7.4780368(3)  & -7.47803657859 \\

 40 & -7.477944869 & -7.47805413551 & -7.4780575(10) & -7.47805422814 \\
  \end{tabular}
  \end{center}
  \end{table}
%
%
  \begin{table}[tbp]
   \caption{Semi-exponential radial basis functions for $N = 28$ with
            optimized exponents. This wave function produces the total
            energy $E$ = -7.47803638005 $a.u.$ for the ground
            $1^2S-$state of the ${}^{\infty}$Li atom. Only one electron
            spin-function $\chi_1 = \alpha \beta \alpha - \beta \alpha
            \alpha$ was used in these calculations.}
     \begin{center}
     \scalebox{0.72}{%
     \begin{tabular}{cccccccccc}
      \hline\hline 
 $N$ & $n_1$ & $n_2$ & $n_3$ & $m_1$ & $m_2$ & $m_3$ & $\alpha$ & $\beta$ & 
 $\gamma$ \\
     \hline
 1  &  0 &  0 &   0 &   0 &   0 &   1 & 0.340570905705403E+01 & 0.293295186982955E+01 & 0.771546188231103E+00 \\

 2  &  0 &  0 &   0 &   1 &   0 &   1 & 0.182145794896563E+01 & 0.329011437023099E+01 & 0.316360558089102E+01 \\

 3  &  0 &  0 &   0 &   1 &   1 &   1 & 0.276859627151255E+01 & 0.297627929771076E+01 & 0.668289237018917E+00 \\

 4  &  0 &  0 &   0 &   2 &   0 &   1 & 0.286729169596989E+01 & 0.300017942427107E+01 & 0.637008666805219E+00 \\

 5  &  0 &  0 &   1 &   0 &   0 &   1 & 0.278208405003856E+01 & 0.275168552095737E+01 & 0.645451709771134E+00 \\

 6  &  0 &  0 &   2 &   0 &   0 &   1 & 0.373836260739635E+01 & 0.339167728695617E+01 & 0.658338285572628E+00 \\

 7  &  0 &  0 &   0 &   0 &   0 &   0 & 0.337994212648889E+01 & 0.324225523017272E+01 & 0.108564332122329E+01 \\

 8  &  1 &  0 &   0 &   0 &   0 &   0 & 0.167110811817316E+01 & 0.338626357605054E+01 & 0.867142603936635E+00 \\

 9  &  0 &  0 &   0 &   0 &   0 &   2 & 0.306529447275211E+01 & 0.299446682705661E+01 & 0.688809932201360E+00 \\

 10 &  1 &  0 &   0 &   0 &   1 &   0 & 0.263305048764256E+01 & 0.242503640146648E+01 & 0.776895925198369E+00 \\

 11 &  0 &  0 &   3 &   0 &   0 &   1 & 0.521025323497647E+01 & 0.361243051429800E+01 & 0.660062464430887E+00 \\

 12 &  1 &  0 &   0 &   0 &   0 &   1 & 0.306893947287392E+01 & 0.208144332232383E+01 & 0.123105191974125E+01 \\

 13 &  0 &  0 &   0 &   0 &   0 &   3 & 0.186312473874842E+01 & 0.289718514568496E+01 & 0.996173732965274E+00 \\

 14 &  0 &  0 &   1 &   1 &   0 &   1 & 0.454831261352940E+01 & 0.219828523532609E+01 & 0.158185067718071E+01 \\

 15 &  0 &  0 &   0 &   3 &   0 &   1 & 0.295559769377747E+01 & 0.327296616683230E+01 & 0.672095727321017E+00 \\

 16 &  0 &  0 &   4 &   0 &   0 &   1 & 0.113723992478698E+02 & 0.508008324231496E+01 & 0.636162427376743E+00 \\

 17 &  0 &  0 &   0 &   2 &   2 &   1 & 0.368595304619283E+01 & 0.278359204737422E+01 & 0.620935860917418E+00 \\

 18 &  0 &  0 &   1 &   1 &   1 &   1 & 0.357466607750891E+01 & 0.343630387369804E+01 & 0.698563354607838E+00 \\

 19 &  0 &  0 &   1 &   2 &   0 &   1 & 0.365095130448352E+01 & 0.317945977760753E+01 & 0.652269908234656E+00 \\

 20 &  0 &  0 &   1 &   3 &   0 &   1 & 0.290005170216145E+01 & 0.289825235917765E+01 & 0.685753232846547E+00 \\

 21 &  2 &  0 &   0 &   0 &   0 &   0 & 0.280340788922781E+01 & 0.225117021991187E+01 & 0.990926897297406E+00 \\

 22 &  1 &  1 &   0 &   0 &   0 &   0 & 0.275272575017714E+01 & 0.276113990248958E+01 & 0.802457922378512E+00 \\

 23 &  1 &  0 &   0 &   0 &   2 &   0 & 0.606571100003296E+01 & 0.326870035708320E+01 & 0.453868977046091E+00 \\

 24 &  1 &  0 &   0 &   1 &   1 &   0 & 0.293743657648302E+01 & 0.226578275529031E+01 & 0.669408905271258E+00 \\

 25 &  0 &  0 &   0 &   0 &   0 &   4 & 0.502952034298549E+01 & 0.276772288773810E+01 & 0.950599883558449E+00 \\

 26 &  0 &  0 &   1 &   0 &   0 &   0 & 0.868992405342042E+01 & 0.374899612568756E+01 & 0.516228416453001E+00 \\

 27 &  0 &  0 &   1 &   0 &   0 &   2 & 0.433237156102395E+01 & 0.278286578531691E+01 & 0.112727477282094E+01 \\

 28 &  0 &  0 &   0 &   1 &   0 &   0 & 0.221163211952812E+01 & 0.498938978770980E+01 & 0.647628325318278E+00 \\
  \end{tabular}}
  \end{center}
  \end{table}
%
%
  \begin{table}[tbp]
   \caption{Semi-exponential radial basis functions for $N = 40$ with
            optimized exponents. This wave function produces the total
            energy $E$ = -7.47805413551 $a.u.$ for the ground
            $1^2S-$state of the ${}^{\infty}$Li atom. Only one electron
            spin-function $\chi_1 = \alpha \beta \alpha - \beta \alpha
            \alpha$ was used in these calculations.}
     \begin{center}
     \scalebox{0.62}{%
     \begin{tabular}{cccccccccc}
      \hline\hline
 $N$ & $n_1$ & $n_2$ & $n_3$ & $m_1$ & $m_2$ & $m_3$ & $\alpha$ & $\beta$ & 
 $\gamma$ \\
     \hline
 1  &  0 &  0 &   0 &   0 &   0 &   1 &  0.309846257882910E+01 &  0.291892609218063E+01 &  0.845633720338414E+00 \\

 2  &  0 &  0 &   0 &   1 &   0 &   1 &  0.190473561336185E+01 &  0.366143431469990E+01 &  0.157818362413849E+01 \\

 3  &  0 &  0 &   0 &   1 &   1 &   1 &  0.278598083293758E+01 &  0.282092658543858E+01 &  0.679821359567157E+00 \\

 4  &  0 &  0 &   0 &   2 &   0 &   1 &  0.288166643409964E+01 &  0.284028201233856E+01 &  0.651027214862940E+00 \\

 5  &  0 &  0 &   1 &   0 &   0 &   1 &  0.294108078850332E+01 &  0.281478615927614E+01 &  0.660626319649503E+00 \\

 6  &  0 &  0 &   2 &   0 &   0 &   1 &  0.354836463958720E+01 &  0.298528920333113E+01 &  0.672210253693381E+00 \\

 7  &  0 &  0 &   0 &   0 &   0 &   0 &  0.287776340439759E+01 &  0.309168157398350E+01 &  0.732676012007879E+00 \\

 8  &  1 &  0 &   0 &   0 &   0 &   0 &  0.126827836395763E+01 &  0.388838466426963E+01 &  0.696918159081991E+00 \\

 9  &  0 &  0 &   0 &   0 &   0 &   2 &  0.310409040480143E+01 &  0.297686082083507E+01 &  0.712624162776256E+00 \\

 10 &  1 &  0 &   0 &   0 &   1 &   0 &  0.304901894182288E+01 &  0.222521841071977E+01 &  0.707931864133508E+00 \\

 11 &  0 &  0 &   3 &   0 &   0 &   1 &  0.410055124696712E+01 &  0.320833161427739E+01 &  0.685502306391946E+00 \\

 12 &  1 &  0 &   0 &   0 &   0 &   1 &  0.319281393757494E+01 &  0.245033872966999E+01 &  0.131890477152115E+01 \\

 13 &  0 &  0 &   0 &   0 &   0 &   3 &  0.212220234455464E+01 &  0.270066283806003E+01 &  0.102980902194010E+01 \\

 14 &  0 &  0 &   1 &   1 &   0 &   1 &  0.385753710808391E+01 &  0.320997710808722E+01 &  0.149747053647776E+01 \\

 15 &  0 &  0 &   0 &   3 &   0 &   1 &  0.290829017069616E+01 &  0.343736715561301E+01 &  0.656999154522055E+00 \\

 16 &  0 &  0 &   4 &   0 &   0 &   1 &  0.762940571197015E+01 &  0.406380788113988E+01 &  0.643367162248368E+00 \\

 17 &  0 &  0 &   0 &   2 &   2 &   1 &  0.337538643016389E+01 &  0.290752238041757E+01 &  0.757319995998642E+00 \\

 18 &  0 &  0 &   1 &   1 &   1 &   1 &  0.338067126368468E+01 &  0.336098394507054E+01 &  0.689553038540639E+00 \\

 19 &  0 &  0 &   1 &   2 &   0 &   1 &  0.376783902187767E+01 &  0.307834289691412E+01 &  0.692428594575284E+00 \\

 20 &  0 &  0 &   1 &   3 &   0 &   1 &  0.302269266874480E+01 &  0.371054683906740E+01 &  0.259164988133888E+01 \\

 21 &  2 &  0 &   0 &   0 &   0 &   0 &  0.174071943396373E+01 &  0.316379880955191E+01 &  0.135613707594630E+01 \\

 22 &  1 &  1 &   0 &   0 &   0 &   0 &  0.196285150420498E+01 &  0.280066377711626E+01 &  0.135193158043399E+01 \\

 23 &  1 &  0 &   0 &   0 &   2 &   0 &  0.308970260481265E+01 &  0.294746895743190E+01 &  0.815386844474244E+00 \\

 24 &  1 &  0 &   0 &   1 &   1 &   0 &  0.251712242684944E+01 &  0.408949306371305E+01 &  0.114312512643339E+01 \\

 25 &  0 &  0 &   0 &   0 &   0 &   4 &  0.321334220677367E+01 &  0.364406360008282E+01 &  0.854918812691407E+00 \\

 26 &  0 &  0 &   1 &   0 &   0 &   0 &  0.690169324304682E+01 &  0.367442145267222E+01 &  0.515986692127293E+00 \\

 27 &  0 &  0 &   1 &   0 &   0 &   2 &  0.719809975272803E+01 &  0.307482033524797E+01 &  0.136078304480511E+01 \\

 28 &  0 &  0 &   0 &   1 &   0 &   0 &  0.363754290732346E+01 &  0.256273214565107E+01 &  0.168423311171287E+01 \\

 29 &  0 &  0 &   0 &   1 &   0 &   2 &  0.356701987266624E+01 &  0.340559537814167E+01 &  0.824959104823405E+00 \\

 30 &  1 &  0 &   1 &   0 &   0 &   0 &  0.244720102707966E+01 &  0.271840294504025E+01 &  0.895838653318275E+00 \\

 31 &  2 &  0 &   0 &   0 &   1 &   0 &  0.315627784159638E+01 &  0.302366248489895E+01 &  0.903526045564038E+00 \\

 32 &  1 &  0 &   0 &   0 &   1 &   1 &  0.325159760628385E+01 &  0.330340729374998E+01 &  0.897876831570133E+00 \\

 33 &  3 &  0 &   0 &   0 &   0 &   0 &  0.338532643878470E+01 &  0.298984569514637E+01 &  0.936601520312646E+00 \\

 34 &  2 &  0 &   0 &   0 &   0 &   1 &  0.348486218895562E+01 &  0.311367325997526E+01 &  0.975295713010522E+00 \\

 35 &  0 &  0 &   5 &   0 &   0 &   1 &  0.575970395842301E+01 &  0.393521634949650E+01 &  0.781294134452016E+00 \\

 36 &  0 &  0 &   0 &   4 &   0 &   1 &  0.352212817969061E+01 &  0.366200645034756E+01 &  0.347518212792864E+01 \\

 37 &  0 &  0 &   1 &   4 &   0 &   1 &  0.384658211093920E+01 &  0.412493146244247E+01 &  0.976334599279947E+00 \\

 38 &  0 &  0 &   0 &   5 &   0 &   1 &  0.359131298628222E+01 &  0.426731681125525E+01 &  0.125620739151171E+01 \\

 39 &  0 &  0 &   2 &   1 &   0 &   1 &  0.294548846533642E+01 &  0.329899178470137E+01 &  0.769541352183112E+00 \\

 40 &  0 &  0 &   2 &   2 &   0 &   1 &  0.348532012166237E+01 &  0.454562709500504E+01 &  0.894468554048998E+00 \\
  \end{tabular}}
  \end{center}
  \end{table}
\end{document}